\documentclass[10pt]{article}
\usepackage{latexsym,amsmath,amssymb,graphics,amscd}
\usepackage{multirow}
\usepackage{booktabs}
\usepackage{tabularx}
\usepackage[hang,footnotesize]{caption}
\usepackage[pdftex]{graphicx}
\usepackage{graphicx}

\addtocounter{footnote}{1}
\makeatletter
\@addtoreset{figure}{section}
\def\thefigure{\thesection.\@arabic\c@figure}
\def\fps@figure{h,t}
\@addtoreset{table}{bsection}
\def\thetable{\thesection.\@arabic\c@table}
\def\fps@table{h, t}
\@addtoreset{equation}{section}

\makeatother

\newtheorem{theorem}{Theorem}[section]

\newtheorem{lemma}[theorem]{Lemma}
\newtheorem{remark}[theorem]{Remark}
\newtheorem{proposition}[theorem]{Proposition}

\newcommand{\bfi}{\bfseries\itshape}
\newsavebox{\savepar}

\makeatletter

\begin{document}

\title{\textbf{Hedging of time discrete auto-regressive stochastic volatility options}}
\author{Joan del Castillo$^{1}$ and Juan-Pablo Ortega$^{2}$}
\date{}
\maketitle

\begin{abstract}
Numerous empirical proofs indicate the adequacy of the time discrete auto-regressive stochastic volatility models introduced by Taylor~\cite{Taylor:Book1, Taylor:Book2} in the description of the log-returns of financial assets. The pricing and hedging of contingent products that use these models for their underlying assets is a non-trivial exercise due to the incomplete nature of the corresponding market. In this paper we apply two volatility estimation techniques available in the literature for these models, namely Kalman filtering and the hierarchical-likelihood approach, in order to implement  various pricing and dynamical hedging strategies. Our study shows that the local risk minimization scheme developed by F\"ollmer, Schweizer, and Sondermann is particularly appropriate in this setup, especially for at and in the money options or for low hedging frequencies.
\end{abstract}

\medskip


\makeatletter
\addtocounter{footnote}{1} \footnotetext{%
Universitat Aut\`onoma de Barcelona. E-08192 Barcelona, Spain. {\texttt{castillo@mat.uab.cat} }}
\addtocounter{footnote}{2} \footnotetext{%
Corresponding author. Centre National de la Recherche Scientifique, D\'{e}partement de Math\'{e}%
matiques de Besan\c{c}on, Universit\'{e} de Franche-Comt\'{e}, UFR des
Sciences et Techniques. 16, route de Gray. F-25030 Besan\c{c}on cedex.
France. {\texttt{Juan-Pablo.Ortega@univ-fcomte.fr} }}
\makeatother

\section{Introduction}

Ever since Black, Merton, and Scholes introduced their celebrated option valuation formula~\cite{black:scholes:paper,merton:bspaper} a great deal of effort has been dedicated to reproduce that result using more realistic stochastic models for the underlying asset. In the discrete time  modeling setup, the GARCH parametric family~\cite{engle:arch, bollerslev:garch} has been a very popular and successful choice for which the option pricing and hedging problem has been profusely studied; see for example~\cite{duan:GARCH:pricing, heston:nandi, Badescu:option:pricing, chorro:guegan:ielpo, ortega:garch:pricing}, and references therein. Even though the GARCH specification can accommodate most stylized features of financial returns like leptokurticity, volatility clustering, and autocorrelation of squared returns, there are mathematical relations between some of their moments that impose undesirable constraints on some of the parameter values. For example, a well-known phenomenon~\cite{carnero:pena:ruiz} has to do with the relation that links the kurtosis of the process with the parameter that accounts for the persistence of the volatility shocks (the sum of the ARCH and the GARCH parameters, bound to be smaller than one in order to ensure stationarity). This relation implies that for highly volatility persistent time series like the ones observed in practice, the ARCH coefficient is automatically forced to be very small, which is in turn incompatible with having sizeable autocorrelation (ACF) for the squares (the ACF of the squares at lag $1$ is linear in this coefficient). This situation aggravates when innovations with fat tails are used in order to better reproduce leptokurticity. 

The structural rigidities associated to the finiteness of the fourth moment are of particular importance when using quadratic hedging methods for there is an imperative need to remain in the category of square summable objects and finite kurtosis is a convenient way to ensure that (see, for example~\cite[Theorem 3.1 (iv)]{ortega:garch:pricing}).

The auto-regressive stochastic volatility (ARSV) models~\cite{Taylor:Book1, Taylor:Book2} that we will carefully describe later on in Section~\ref{Auto-regressive stochastic volatility (ARSV) models} are a parametric family designed in part to overcome the difficulties that we just explained. The defining feature of these models is the fact that the volatility (or a function of it) is determined by an autoregressive process that, unlike the GARCH situation, is exclusively driven by past volatility values and by designated innovations that are in principle independent from those driving the returns. The use of additional innovations introduces in the model a huge structural leeway because, this time around, the same constraint that ensures stationarity guarantees the existence of finite moments of arbitrary order which is of much value in the context of option pricing and hedging.

The modifications in the ARSV prescription that allow an enriching of the dynamics come with a price that has to do with the fact that conditional volatilities are not determined, as it was the case in the GARCH situation, by the price process. This causes added difficulties at the time of volatility and parameter estimation; in particular, a conditional likelihood cannot be written down in this context (see later on in Section~\ref{Volatility and model estimation}). This also has a serious impact when these models are used in derivatives pricing and hedging because the existence of additional noise sources driving the dynamics makes more acute the incomplete character of the corresponding market; this makes more complex the pricing and hedging of contingent products.

\emph{The main goal of this paper is showing that appropriate implementations of the local risk minimization scheme developed by F\"ollmer, Schweizer, and Sondermann~\cite{foellmer:sondermann, foellmer:schweizer, schweizer:2001} tailored to the ARSV setup provide an appropriate tool for pricing and hedging contingent products that use these models to characterize their underlying assets}. The realization of this hedging scheme in the ARSV context requires two main tools:
\begin{itemize}
\item {\bf Use of a volatility estimation technique:} as we already said, volatility for these models is not determined by the price process and hence becomes a genuinely hidden variable that needs to be estimated separately. In our work we will use a {\bf Kalman filtering} approach advocated by~\cite{harvey:ruiz:shephard} and the so called {\bf hierarchical likelihood} ($h$-likelihood) strategy~\cite{lee:nelder:1, lee:nelder:2, delcastillo:lee:1, delcastillo:lee:2} which, roughly speaking, consists of carrying out a likelihood estimation while considering the volatilities as unobserved parameters that are part of the optimization problem. Even though both methods are adequate  volatility estimation techniques, the $h$-likelihood technique has a much wider range of applicability for it is not subjected to the rigidity of the state space representation necessary for Kalman and hence can be used for  stochastic volatility models with  complex link functions or when innovations are non-Gaussian. 
\item {\bf Use of a pricing kernel:} this term refers to a probability measure equivalent to the physical one with respect to which the discounted prices are martingales. The expectation of the discounted payoff of a derivative product with respect to any of these equivalent measures yields an arbitrage free price for it. In our work we will also use these martingale measures in order to devise hedging strategies via local risk minimization. This is admittedly just an approximation of the optimal local risk minimization scheme that should be carried out with respect to the physical measure that actually quantifies the local risk. However there are good reasons that advise the use of these equivalent martingale measures mainly having to do with numerical computability and interpretation of the resulting value processes as arbitrage free prices. These arguments are analyzed in Section~\ref{Local risk minimization with respect to a martingale measure}.
\end{itemize}
Regarding the last point, there are two equivalent   martingale measures that we will be using. The first one is inspired by the {\it Extended Girsanov Principle} introduced in~\cite{elliott:madan:mcmm}. This principle allows the construction of a martingale measure in a general setup under which the process behaves as its ``martingale component'' used to do under the physical probability; this is the reason why this measure is sometimes referred to as the {\bf mean-correcting martingale measure}, denomination that we will adopt. This construction has been widely used in the GARCH context (see~\cite{Badescu:option:pricing, ortega:garch:pricing} and references therein)  where it admits a particularly simple and numerically efficient expression in terms of the probability density function of the model innovations. As we will see, in the ARSV case, this feature is not anymore available and we will hence work with the expression coming from the predictable situation that even though in the ARSV case produces a measure that does not satisfy the Extended Girsanov Principle, it is still a martingale measure.
Secondly, in Theorem~\ref{introduction minimal martingale measure}  we construct the so called {\bf minimal martingale measure} $Q_{{\rm min}}$ in the ARSV setup; the importance in our context of this measure is given by the fact that  the value process of the local risk-minimizing strategy {\it with respect to the physical measure} for a derivative product coincides with its arbitrage free price when using $Q_{{\rm min}}$ as a pricing kernel. A concern with $Q_{{\rm min}}$ in the ARSV setup  is that {\it this measure is in general signed}; fortunately, the occurrence of negative Radon-Nikodym derivatives is extremely rare for the usual parameter values that one encounters in financial time series. Consequently, the bias introduced by censoring paths that yield negative Radon-Nikodym derivatives and using  $Q _{{\rm min}} $ as a well-defined positive measure is hardly noticeable. A point that is worth emphasizing is that even though the value processes obtained when carrying out local risk minimization with respect to the physical and the minimal martingale measures are identical, the hedges are in general {\it not} the same and consequently so are the hedging errors; this difference is studied in Proposition~\ref{hedges under change of measure}.

A general remark about local risk minimization that supports its choice in applications is its {\bf adaptability to different hedging frequencies}. Most sensitivity based (delta) hedging methods for time series type underlyings are constructed by discretizing a continuous time hedging argument; consequently, when the hedging frequency diminishes, the use of such techniques loses pertinence. As we explain in Section~\ref{Local risk minimization and changes in the hedging frequency}, the local risk minimization hedging scheme can be adapted to prescribed changes in the hedging frequency, which regularly happens in real life applications.

The paper is organized in four sections. Section~\ref{Auto-regressive stochastic volatility (ARSV) models} contains a brief introduction to ARSV models and their dynamical features of interest to our study; this section contains two subsections that explain the volatility estimation techniques that we will be using and the martingale measures that we mentioned above. The details about the implementation of the local risk minimization strategy are contained in Section~\ref{Local risk minimization for ARSV options}. Section~\ref{Empirical study} contains a numerical study where we carry out a comparison between the
hedging performances obtained by implementing the local risk minimization scheme using the different volatility estimation techniques and the martingale measures introduced in Section~\ref{Auto-regressive stochastic volatility (ARSV) models}. In order to enrich the discussion, we have added to the study two standard sensitivity based hedging methods that provide good results in other contexts, namely Black-Scholes~\cite{black:scholes:paper} and an adaptation of Duan's static hedge~\cite{duan:GARCH:pricing} to the ARSV context. As a summary of the results obtained in that section it can be said that {\it local risk minimization outperforms sensitivity based hedging methods, especially when maturities are long and the hedging frequencies are low}; this could be expected due to the adaptability feature of local risk minimization with respect to modifications in this variable. As to the different measures proposed, {\it it is the minimal martingale measure that yields the best hedging performances when implemented using a Kalman based estimation of volatilities, in the case of short and medium maturities and $h$-likelihood for longer maturities and low hedging frequencies}.

\medskip

\noindent {\bf Conventions and notations:} The proofs of all the results in the paper are contained in the appendix in Section~\ref{Appendix}.
Given a filtered probability space $(\Omega, P, \mathcal{F}, \{ \mathcal{F} _t\}_{t \in \mathbb{N}})$ and $X,Y $ two random variables, we will denote by $E _t[X]:=E[X| \mathcal{F}  _t]$ the conditional expectation, ${\rm cov} _t(X,Y):= {\rm cov}(X,Y| \mathcal{F} _t):=E_t[XY]-E_t[X]E _t[Y]$ the conditional covariance, and by ${\rm var} _t(X):=E_t[X ^2]-E _t[X] ^2 $ the conditional variance. A discrete-time stochastic process $\{X _t\}_{t \in  \mathbb{N}} $ is predictable when $X _t$ is $\mathcal{F} _{t-1} $-measurable, for any $t \in \mathbb{N} $.

\medskip

\noindent {\bf  Acknowledgments:} we thank Alexandru Badescu for insightful comments about our work that have significantly improved its overall quality.

\section{Auto-regressive stochastic volatility (ARSV) models}
\label{Auto-regressive stochastic volatility (ARSV) models}

The auto-regressive stochastic volatility (ARSV) model has been introduced in~\cite{Taylor:Book1} with the objective of capturing some of the most common stylized features observed in the excess returns of financial time series: volatility clustering, excess kurtosis, and autodecorrelation in the presence of dependence; this last feature can be visualized by noticing that  financial log-returns time series exhibit autocorrelation at, say lag $1$, close to zero while the autocorrelation of the squared returns is significantly not null. Let $S _t $ be the price at time $t$ of the asset under consideration, $r$ the risk-free interest rate, and $y _t:=\log\left(S _t/S_{t-1} \right) $ the associated log-return. In this paper we will be considering the so-called {\bf standard ARSV} model which is given by the prescription
\begin{equation}
\label{arsv model}
\left\{
\begin{array}{rcl}
y _t &= &r+ \sigma _t \epsilon _t, \qquad \{ \epsilon _t\} \sim {\rm IIDN}(0,1)\\
b _t&= & \gamma+ \phi b_{t-1}+ w _t,  \qquad \{ w _t\} \sim {\rm IIDN}(0,\sigma _w^2)
\end{array}
\right.
\end{equation}
where $b _t:= \log (\sigma_t^2)$, $\gamma  $ is a real parameter, and $\phi \in (-1,1) $. Notice that in this model, the volatility process $\{ \sigma _t\} $ is a non-traded stochastic latent variable that, unlike the situation in GARCH like models~\cite{engle:arch, bollerslev:garch} is not a predictable process that can be written as a function of previous returns and volatilities.

It is easy to prove that the unique stationary returns process induced by~(\ref{arsv model}) available in the presence of the constraint $\phi \in (-1,1) $ is a white noise (the returns have no autocorrelation) with finite moments of arbitrary order.  In particular, the marginal variance and kurtosis are given by
\begin{equation}
\label{variance and kurtosis}
{\rm var}(y _t)= {\rm E} [ \sigma _t^2]= \exp\left[ \frac{\gamma}{1- \phi}+ \frac{1}{2} \sigma _b ^2\right], \qquad {\rm kurtosis}\,(y _t)= 3\exp \left( \sigma _b ^2 \right)
\end{equation}
Moreover, let $\sigma _b ^2 $ be the marginal variance of the stationary process $\{ b _t \} $, that is,
\begin{equation*}
\sigma _b ^2= \frac{\sigma_w ^2}{1- \phi^2}.
\end{equation*}
It can be shown~\cite{Taylor:Book1} that whenever $\sigma _b ^2 $ is small and/or $\phi$ is close to one then the autocorrelation $\gamma (h) $  of the squared returns at lag  $h$ can be approximated by
\begin{equation*}
\gamma (h)\simeq \frac{\exp(\sigma_b ^2)-1}{3\exp(\sigma_b ^2)-1}\phi ^h.
\end{equation*}
The existence of finite moments is particularly convenient in the context of quadratic hedging methods. For example, in Theorem 3.1 of~\cite{ortega:garch:pricing} it is shown that the finiteness of kurtosis is a sufficient condition for the availability of adequate integrability conditions necessary to carry out pricing and hedging via local risk minimization.

Let $(\Omega, P) $  be the probability space where the model~(\ref{arsv model}) has been formulated and let $\mathcal{F}_t $ be the information set generated by the observables $\{S _0, S _1, \ldots, S_t \} $. This statement can be mathematically coded by setting $\mathcal{F}_t= \sigma\left(S _0, S _1, \ldots, S_t \right) $, where $\sigma\left(S _0, S _1, \ldots, S_t \right) $ is the sigma algebra generated by the prices $\{S _0, S _1, \ldots, S_t\} $. As several equivalent probability measures will appear in our discussion, we will refer to $P$ as the {\bf physical} or {\bf historical} probability measure. 

Unless we indicate otherwise, in the rest of the discussion we will not assume that the innovations $\{ \epsilon _t\}  $  are Gaussian.
We will also need the following marginal and {\bf conditional cumulant functions} of $\epsilon _t  $ with respect to the filtration $\mathcal{F}=\{ \mathcal{F} _t\}_{t \in \{0, \ldots, T\}} $. 
\begin{eqnarray*}
L_{\epsilon _t}^P(z)&=&\log E^P\left[e^{z \epsilon _t}\right], \qquad z \in \mathbb{R},\\
K_{\epsilon _t}^P(u)&=&\log E^P\left[e^{u \epsilon _t}| {\cal F}_{t-1}\right], \qquad \text{with } u  \mbox{ a random variable}. \quad
\end{eqnarray*}
If the innovations $\{ \epsilon _t\}  $  are Gaussian, we obviously have $L_{\epsilon _t}^P(z)=z ^2/2 $. When the random variables $u$ and $\epsilon_t $ are $P$-independent, the following relation between $L_{\epsilon _t}^P $  and $K_{\epsilon _t}^P $ holds.
\begin{lemma}
\label{relation cumulant functions}
Let $u:( \Omega,P) \rightarrow \mathbb{R}$ be a random variable independent of $\epsilon _t $ and $\mathcal{F}=\{ \mathcal{F} _t\}_{t \in \{0, \ldots, T\}} $ the filtration introduced above. Then
\begin{equation}
\label{equality with Ks}
K_{\epsilon _t}^P(u)=\log E^P\left[e^{L^P_{\epsilon _t}(u)}\mid {\cal F}_{t-1}\right].
\end{equation}
\end{lemma}
The cumulant functions that we just introduced are very useful at the time of writing down the conditional means and variances of price changes with respect to the physical probability; these quantities will show up frequently in our developments later on. Before we provide these expressions we introduce the following notation for discounted prices:
\begin{equation*}
\widetilde{S _t}:=S _te^{-rt},
\end{equation*}
where $r$ denotes the risk-free interest rate. The tilde will be used in general for discounted processes. Using this notation and Lemma~\ref{relation cumulant functions}, a straightforward computation using the independence at any given time step $t$ of the innovations $\{\epsilon_t \}$ and the volatility process $\{ \sigma _t \} $ defined by ~(\ref{arsv model}), shows that
\begin{eqnarray}
E^P\left[\widetilde{S _t}-\widetilde{S} _{t-1}\mid \mathcal{F}_{t-1}\right] &=&\widetilde{S} _{t-1} \left(e^{K_{\epsilon _t}^P (\sigma _t)}-1\right)=\widetilde{S} _{t-1}E^P\left[e^{L^P_{\epsilon _t}(\sigma _t)}\mid {\cal F}_{t-1}\right],\label{expected increase 1}\\
{\rm var}\left[\widetilde{S _t}-\widetilde{S} _{t-1}\mid \mathcal{F}_{t-1}\right]  &= & \widetilde{S} _{t-1} ^2 \left[e^{K_{\epsilon _t}^P(2\sigma _t)}-e^{2K_{\epsilon _t}^P(\sigma _t)}\right].\label{expected increase 2}
\end{eqnarray}

\subsection{Volatility and model estimation}
\label{Volatility and model estimation}

 The main complication in the estimation of ARSV models is due to the fact that $\mathcal{F}_t $ does not determine the random variable $\sigma _t $ and hence makes impossible the writing of a likelihood function in a traditional sense. Many procedures have been developed over the years to go around this difficulty based on different techniques: moment matching~\cite{Taylor:Book1}, generalized method of moments~\cite{melino:turnbull, jacquier:polson:rossi, andersen:sorensen}, combinations of quasi-maximum likelihood with the Kalman filter~\cite{harvey:ruiz:shephard}, simulated maximum likelihood~\cite{danielsson;simulated:ml}, MCMC~\cite{jacquier:polson:rossi} and, more recently, hierarchical-likelihood~\cite{lee:nelder:1, lee:nelder:2, delcastillo:lee:1, delcastillo:lee:2} (abbreviated in what follows as h-likelihood). An excellent overview of some of these methods is provided in Chapter 11 of the monograph~\cite{Taylor:Book2}.

In this paper we will focus on the Kalman and h-likelihood approaches since both are based on numerically efficient estimations of the volatility, a point of much importance in our developments. 

\medskip

\noindent {\bf State space representation and Kalman filtering.} This approach is advocated by~\cite{harvey:ruiz:shephard} and consists of writing down the model~(\ref{arsv model}) using a state space representation by setting
\begin{equation*}
l _t=\log \left(|y _t-r|\right), \qquad L _t=\log \left(\sigma _t\right).
\end{equation*}
We assume in this paragraph that the innovations $\{\epsilon _t \}$ are Gaussian. Using these variables,~(\ref{arsv model}) can be rephrased as the observation and state equations:
\begin{equation}
\label{arsv model kalman form}
\left\{
\begin{array}{rcl}
l _t &= &L _t+ \xi_t, \qquad \xi _t=\log \left(| \epsilon _t|\right),\\
(L _t - \alpha _K)&= &\phi (L _{t-1} - \alpha _K)+ \eta_t,  \qquad  \eta_t= \frac{1}{2}w _t, 
\end{array}
\right.
\end{equation}
where $ \alpha _K= E [L _t]= \frac{1}{2}E[b _t]= \frac{\gamma}{2(1- \phi)} $. Kalman filtering cannot be directly applied to~(\ref{arsv model kalman form}) due to the non-Gaussian character of the innovations $\xi _t=\log \left(| \epsilon _t|\right) $ hence,  the approximation in~\cite{harvey:ruiz:shephard} consists of using this procedure by thinking of $\xi _t $ as a Gaussian variable with mean $\mu _{\xi}:=E[\xi _t]=-0.63518 $  and variance $\sigma^2_{\xi }=  {\rm Var}[\xi_t] = \pi ^2/8$. The Kalman filter provides an algorithm to iteratively produce one-step ahead linear forecasts $l _{t,1}:=P[l _t| \mathcal{F} _{t-1}]$; given that  $l_t= \log (\sigma _t)+\log \left(| \epsilon _t|\right)  $, these forecasts can be used to produce predictable estimations $\sigma_t^k  $ of the volatilities $\sigma _t $ by setting $\log \left(\sigma_t^k\right):= l_{t-1,1}- \mu_\xi $ and hence
\begin{equation}
\label{kalman vola}
\sigma_t^k=\exp (l_{t-1,1}- \mu_\xi).
\end{equation}

\medskip

\noindent {\bf The $h$-likelihood approach.} This method~\cite{lee:nelder:1, lee:nelder:2, delcastillo:lee:1, delcastillo:lee:2} consists of carrying out a likelihood estimation while considering the volatilities as unobserved parameters that are part of the optimization problem. More specifically, let $\alpha:=(\gamma, \phi, \sigma _w) $ be the parameters vector, $b=(b _1, \ldots, b _T)$ the vector that contains the unobserved  $b _t:=\log (\sigma_t^2) $, and let $z _t:= y _t- r  $; if $f(z _t|b _t)$ and $f(b| \alpha ) $ denote the corresponding probability density functions, the associated $h$-(log)likelihood is defined as
\begin{eqnarray}
h(z;b, \alpha) &= &\sum_{t=1}^T\log f(z _t|b _t)+\log f(b| \alpha )\notag\\
	&= &- \frac{1}{2}\sum_{t=1}^T \left(z _t ^2\exp (-b _t) + b _t+ \frac{1}{\sigma _w ^2}(b _t- \gamma- \phi b _{t-1})^2+\log \sigma _w ^2\right).\label{h-likelihood f}
\end{eqnarray}
The references listed above provide numerically efficient procedures to maximize~(\ref{h-likelihood f}) with respect to the variables $b$ and $\alpha$ once the sample $z$ has been specified. An idea of much importance in our context is that, once the coefficients $\alpha $ have been estimated, this procedure has a natural one-step ahead forecast of the volatility associated, similar to the one obtained using the Kalman filter. The technique operates using the variables $b _t:=\log (\sigma_t^2) $ alternating prediction and updating; more specifically, we initialize the algorithm using the mean of the unique stationary solution of~(\ref{arsv model}), that is, $b _0:= \gamma/(1- \phi) $. We use the second equation in~(\ref{arsv model}) to produce a linear prediction $ b_{1p} $ of $b _1 $ based on the value of $b _0$, namely $b_{1p}= \gamma+ \phi b _0 $. Once the quote $z _1 $ is available, the h-likelihood approach can be used to update this forecast to the value $b_{1u} $ by solving the optimization problem
\begin{equation*}
b_{1u}=\mathop{\rm arg\, min}_{b _1\in \mathbb{R}} \left(z _1 ^2\exp (-b _1) + b _1+ \frac{1}{\sigma _w ^2}(b _1- \gamma- \phi b _{0})^2 \right).
\end{equation*}
This value can be used in its turn to produce a forecast $b_{2p} $ for $b _2 $ by setting $b_{2p}= \gamma+ \phi b _{1u} $. If we continue this pattern we obtain a succession of prediction/updating steps given by the expressions
\begin{eqnarray*}
b_{tp}&=& \gamma+ \phi b _{(t-1)u},\\
b_{tu}&=&\mathop{\rm arg\, min}_{b _t\in \mathbb{R}} \left(z _t ^2\exp (-b _t) + b _t+ \frac{1}{\sigma _w ^2}(b _t- \gamma- \phi b _{(t-1)u})^2 \right).
\end{eqnarray*}
These forecasts can be used to  produce predictable estimations $\sigma_t^h  $ of the volatilities $\sigma _t $ by setting
\begin{equation*}
\sigma_t^h:=\exp \left(\frac{1}{2} b _{tp}\right).
\end{equation*}

\begin{remark}
\normalfont
Even though both the Kalman and the $h$-likelihood methods constitute adequate estimation and volatility forecasting methods, the $h$-likelihood technique has a much wider range of applicability for it is not subjected to the rigidity of the state space representation and hence can be generalized to stochastic volatility models with more complex link functions than the second expression in~(\ref{arsv model}), to situations where the innovations are non-Gaussian, or there is a dependence between $\{ \epsilon _t\} $ and  $\{ w _t\} $. 
\end{remark}

\subsection{Equivalent martingale  measures for stochastic volatility models}
\label{Equivalent martingale and quasi-martingale measures for stochastic volatility models}

Any technique for the pricing and hedging of options based on no-arbitrage theory requires the use of an equivalent measure for the probability space used for the modeling of the underlying asset under which its discounted prices are martingales with respect to the filtration generated by the observables. Measures with this property are usually referred to as risk-neutral or simply martingale measures and the Radon-Nikodym derivative that links this measure with the historical or physical one is called a pricing kernel. 

A number of martingale measures have been formulated in the literature in the context of GARCH-like time discrete processes with predictable conditional volatility; see~\cite{Badescu:option:pricing} for a good comparative account of many of them. Those constructions do not generalize to the SV context  mainly due to the fact that the volatility process $\{\sigma _t \} $ is not uniquely determined by the price process $ \{S _t\} $ and hence it is not predictable with respect to the filtration $ \mathcal{F}=\{\mathcal{F}_t \}$ generated by $ \{S _t\} $.

In this piece of work we will explore two solutions to this problem that, when applied to option pricing and hedging, yields a good combination of theoretical computability and numerical efficiency. The first one has to do with the so called minimal martingale measure and the second approach is based on an approximation inspired in the Extended Girsanov Principle~\cite{elliott:madan:mcmm}. 

\medskip

\noindent {\bf The minimal martingale measure.} As we will see in the next section, this measure is particularly convenient when using local risk minimization with respect to the physical measure~\cite{foellmer:sondermann, foellmer:schweizer, foellmer:schied:book} as a pricing/hedging technique, for it provides the necessary tools to interpret the associated value process   as an arbitrage free price for the contingent product under consideration. 

The minimal martingale measure $Q_{{\rm min}}$ is a measure equivalent to $P $ defined by the following property:  every $P$-martingale $M \in L ^2(\Omega,P)$ that is strongly orthogonal to the discounted price process $\widetilde{S}$, is also a $Q_{{\rm min}}$-martingale. The following result spells out the specific form that $Q _{{\rm min}} $ takes when it comes to the ARSV models.

\begin{theorem}
\label{introduction minimal martingale measure}
Consider the price process $S=\{S _0, S _1, \ldots, S_T\} $ associated to the ARSV model given by the expression~(\ref{arsv model}). In this setup, the minimal martingale measure is determined by the Radon-Nikodym derivative $dQ _{{\rm min}}/ d P   $ that is obtained by evaluating at time $T$ the $P$-martingale $\{Z _t\}_{t \in \{1, \ldots, T\}} $ defined by
\begin{equation}
\label{mmm sv expression}
Z _t:=\prod_{k=1}^t \left(1+ \frac{\left(e^{K_{\epsilon_t}^P(\sigma _k)}-1\right)\left(e^{\sigma _k \epsilon _k}-e^{K_{\epsilon_t}^P(\sigma _k)}\right)}{e^{2K_{\epsilon_t}^P(\sigma _k)}-e^{K_{\epsilon_t}^P(2\sigma _k)}}\right).
\end{equation}
\end{theorem}

\noindent \noindent\textbf{Proof.\ \ } By Corollaries 10.28 and 10.29 and Theorem 10.30 in~\cite{foellmer:schied:book}, the minimal martingale measure, when it exists, is unique and is determined by the Radon-Nikodym derivative obtained by evaluating at $T$ the $P$-martingale
\begin{equation}
\label{mmm general expression}
Z _t:=\prod_{k=1}^t 1+ \lambda _k\left(\widetilde{Y} _k- \widetilde{Y}_{k-1}\right), \quad \mbox{where} \quad \lambda _k:=- \frac{E^P_{k-1}\left[\widetilde{S} _k- \widetilde{S}_{k-1}\right]}{{\rm var}_{k-1} \left(\widetilde{S} _k- \widetilde{S}_{k-1}\right)},
\end{equation}
and $\widetilde{Y} _k $ is the martingale part in the Doob decomposition of $\widetilde{S} _k $ with respect to $P$. Hence we have
\begin{equation}
\label{doob part sv}
\widetilde{Y} _k- \widetilde{Y}_{k-1}=\widetilde{S} _k- E^P_{k-1}\left[\widetilde{S}_{k}\right]=\widetilde{S} _{k-1} \left(e^{\sigma _k \epsilon _k}-e^{K^P_{\epsilon _k}(\sigma _k)}\right).
\end{equation}
The equality~(\ref{mmm sv expression}) follows from~(\ref{mmm general expression}),~(\ref{doob part sv}),~(\ref{expected increase 1}), and~(\ref{expected increase 2}). \quad $\blacksquare$

\begin{remark}
\label{minimal is signed}
\normalfont
The measure $Q _{{\rm min}} $  obtained by using~(\ref{mmm sv expression}) {\it is in general signed}. Indeed, as the random variable $ \sigma_k \epsilon_k $  is $\mathcal{F} _k $-adapted and $K_{\epsilon _k}^P(\sigma _k) $ is $\mathcal{F} _k $-predictable, the term $\left(e^{\sigma _k \epsilon _k}-e^{K^P_{\epsilon _k}(\sigma _k)}\right) $ can take arbitrarily negative values that can force $Z _t $ to become negative. We will see in our numerical experiments  that even though this is in general the case, negative occurrences are extremely unlikely for the usual parameter values that one encounters in financial time series. Consequently, the bias introduced by censoring paths that yield negative Radon-Nikodym derivatives and using  $Q _{{\rm min}} $ as a well-defined positive measure is not noticeable.
\end{remark}

\begin{remark}
\normalfont
The quantity $K_{\epsilon _t}^P(\sigma _t)=\log E^P\left[e^{\sigma_t \epsilon _t}| {\cal F}_{t-1}\right]$ in~(\ref{mmm sv expression}) is in general extremely difficult to compute. Lemma~\ref{relation cumulant functions} allows us to rewrite it as a conditional expectation of a function that depends only on $\sigma _t $ and hence it is reasonable to use the estimations $\sigma_t^k $ and $\sigma_t^h $ introduced in Section~\ref{Volatility and model estimation} to approximate $K_{\epsilon _t}^P(\sigma _t) $. For example, if the innovations $\{ \epsilon _t\}  $  are Gaussian and we hence have $L_{\epsilon _t}^P(z)=z ^2/2 $, we will approximate
\begin{equation*}
K_{\epsilon _t}^P(\sigma _t)=\log E^P_{t-1}\left[e^{\frac{\sigma_t^2}{2}}\right]
\end{equation*}
by $(\sigma_t^k) ^2/2  $ or $(\sigma_t^h) ^2/2  $ depending on the technique (Kalman or $h$-likelihood, respectively) used to estimate the conditional volatility.
\end{remark}

\medskip

\noindent {\bf The Extended Girsanov Principle and the mean correcting martingale measure.} This construction has been introduced in~\cite{elliott:madan:mcmm} as an extension in discrete time and for multivariate processes of the classical Girsanov Theorem. This measure is designed so that when the process is considered with respect to it, its dynamical behavior coincides with that of its martingale component under the original historical probability measure. These martingale measures are widely used in the GARCH context (see for example~\cite{Badescu:option:pricing, ortega:garch:pricing} and references therein).

In this paragraph we will work with models that are slightly more general than~(\ref{arsv model}) in the sense that we will allow for predictable trend terms and generalized innovations, that is, our ARSV model will take the form
\begin{equation}
\label{arsv model general}
\left\{
\begin{array}{rcl}
y _t &= &m _t+ \sigma _t \epsilon _t, \qquad \{ \epsilon _t\} \sim {\rm IID}(0,1)\\
b _t&= & \gamma+ \phi b_{t-1}+ w _t,  \qquad \{ w _t\} \sim {\rm IID}(0,\sigma _w^2)
\end{array}
\right.
\end{equation}
with $m _t  $ an $\mathcal{F}_{t-1} $-measurable random variable. 
\begin{theorem}
\label{mcmm arsv expression}
Given the model specified by~(\ref{arsv model general}), let $f_{\sigma _t \epsilon _t} ^P$ be the conditional probability density function under $P$ of the random variable $\sigma _t \epsilon _t $ given $\mathcal{F} _{t-1}$ and let $\{ N _t \} $ be the stochastic discount factors defined by:
\begin{equation*}
N _t:=\frac{f^P_{\sigma_t \epsilon _t}(\sigma _t\epsilon _t+ m _t -r+\log E^P_{t-1}\left[e^{\sigma _t\epsilon _t}\right])}{f^P_{\sigma_t \epsilon _t}(\sigma _t\epsilon _t)}.
\end{equation*}
The process $Z _t:=\prod_{k=1}^t N _k $,  is  a $(\mathcal{F}_t,P)$-martingale such that  $E ^P[Z _t]=1 $ and $Z _T$ defines in the ARSV setup   the martingale measure $Q _{{\rm EGP}}$ associated to Extended Girsanov Principle   via the identity  $Z _T= d Q_{{\rm EGP}}/d P$.    
\end{theorem}

\begin{remark}
\normalfont
As $\sigma_t $ and $\epsilon _t $ are independent processes, Rohatgi's formula yields:
\begin{equation*}
f^P_{\sigma_t \epsilon _t} (z)=\int_{-\infty}^\infty \frac{1}{|x|}f^P_{\sigma_t} (x) f^P_{\epsilon _t} \left(\frac{z}{x}\right) d x.
\end{equation*}
Unfortunately, there is no closed form expression for this integral even in the Gaussian setup.  Though {\it ad hoc} numerical have been developed to treat it (see for instance~\cite{glenetal:product:distributions}), the computation of the stochastic discount factors $N _t $  for each time step $t$ may prove to be a computationally  heavy task.
\end{remark}

The point that we just raised in the remark leads us to introduce yet another martingale measure that mimics the good analytical properties of the Extended Girsanov Principle in the GARCH case but that, in the ARSV setup, only shares approximately its other defining features. In order to present this construction we define the {\bf market price of risk} process:
\begin{equation}
\label{market prices of risk}
\rho_t= \frac{m _t+K_{\epsilon _t}^P(\sigma _t)-r}{\sigma _t},
\end{equation}
to which we can associate what we will call the {\bf stochastic discount factor}:
\begin{equation*}
N _t(\epsilon _t,  \rho _t ) = \frac{f_t ^P(\epsilon_t+ \rho _t)}{f_t ^P(\epsilon _t)}, 
\end{equation*}
where $f_t ^P $ is the conditional probability density function of the innovations $\{ \epsilon _t\} $ with respect to the measure $P$ given $ \mathcal{F} _{t-1}$.
\begin{theorem}
\label{quasi mean correcting martingale measure}
Using the notation that we just introduced, we have that
\begin{description}
\item [(i)]The process $Z _t:=\prod_{k=1}^t N _k $,  is  a $(\mathcal{F}_t,P)$-martingale such that  $E ^P[Z _t]=1 $.
\item [(ii)] $Z _T$ defines an equivalent measure $Q _{{\rm mc}}$ such that $Z _T= d Q_{{\rm mc}}/d P$ under which the discounted price process $\left\{S _0, S _1, \ldots, S _T\right\}$ determined by~(\ref{arsv model general}) is a martingale. By an abuse of language we will refer to $Q_{{\rm mc}} $ as the ARSV {\bf  mean correcting martingale measure}.

\end{description}
\end{theorem}

\section{Local risk minimization for ARSV options}
\label{Local risk minimization for ARSV options}

In this section we use the local risk minimization pricing/hedging technique developed by F\"ollmer, Schweizer, and Sondermann (see~\cite{foellmer:sondermann, foellmer:schweizer, schweizer:2001}, and references therein) as well as the different measures and volatility estimation techniques introduced in the previous section to come up with prices and hedging ratios for European options that have an ARSV process as model for the underlying asset.

As we will see, this technique provides simultaneous expressions for prices and hedging ratios that, even though require Monte Carlo computations in most cases, admit convenient interpretations based on the notion of hedging error minimization that can be adapted to various models for the underlying asset. This hedging approach has been studied in~\cite{ortega:garch:pricing} in the context of GARCH models with Gaussian innovations and in~\cite{Badescu:Ortega} for more general innovations. Additionally, the technique can be tuned in order to accommodate different prescribed hedging frequencies and hence adapts very well to realistic situations encountered by practitioners. In the negative side, like any other quadratic method, local risk minimization penalizes equally shortfall and windfall hedging errors, which may sometimes lead to inadequate hedging decisions.

\subsection{Generalized trading strategies and local risk minimization}

We now briefly review the necessary concepts on pricing by local risk minimization that we will needed in the sequel. The reader is encouraged to check with Chapter 10 of the excellent monograph~\cite{foellmer:schied:book} for a self-contained and comprehensive presentation of the subject.

As we have done so far, we will denote by $S _t $ the price of the underlying asset at time $t$. The symbol $r _t $ denotes the continuously composed  risk-free interest rate paid on the currency of the underlying in the period that goes from time $t-1 $ to $t$; we will assume that $\{ r _t\} $  is a predictable process. Denote by 
\begin{equation*}
R _t:=\sum_{j=0}^t r _j.
\end{equation*}
The price at time $t$ of the riskless asset $S ^0 $ such that  $S _0 ^0=1 $, is given by $S _t ^0= {\rm e}^{R _t}$.

Let now $H(S _T)$ be a European contingent claim that depends on the terminal value of the risky asset  $S _t $. In the context of an incomplete market, it will be in general impossible to replicate the payoff $H$ by using a self-financing portfolio. Therefore, we introduce the notion of {\bf generalized trading strategy}, in which the possibility of additional investment in the riskless asset throughout the trading periods up to expiry time $T$ is allowed. All the following statements are made with respect to a fixed filtered probability space $(\Omega, P, \mathcal{F}, \{ \mathcal{F} _t\}_{t \in\{0, \ldots, T\}})$:
\begin{itemize}
\item A {\bfi  generalized trading strategy} is a pair of stochastic processes $(\xi^0, \xi)$ such that  $\{\xi^0_t\}_{t \in \{0, \ldots,T\}}$    is adapted and $\{\xi_t\} _{t \in \{1, \ldots,T\}}$ is predictable. The associated {\bfi  value process} $V $ of $(\xi^0, \xi) $ is defined as
\begin{equation*}
V _0:= \xi_0^0, \quad \mbox{ and  } \quad V _t:= \xi_t^0 \cdot S ^0_t+ \xi_t\cdot S _t, \quad t\geq 1.
\end{equation*}
\item The {\bfi  gains process} $G$ of the generalized trading strategy $(\xi^0, \xi)$ is given by
\begin{equation*}
G _0:=0 \quad \mbox{and } \quad G _t:=\sum_{k=1}^t \xi_k\cdot (S _k-S_{k-1}), \quad t=0, \ldots ,T.
\end{equation*}
\item The {\bfi  cost process} $C$ is defined by the difference
\begin{equation*}
C _t:=V _t - G _t, \quad t=0, \ldots, T.
\end{equation*}
\end{itemize}
These processes have discounted versions $\widetilde{V} _t $, $\widetilde{G} _t $, and $\widetilde{C} _t $ defined as:
\begin{equation*}
\widetilde{V }_t:= V _t {\rm e}^{-R _t}, \qquad \widetilde{G} _t:=\sum_{k=1}^t \xi_k\cdot (\widetilde{S} _k-\widetilde{S}_{k-1}), \quad \mbox{and} \quad \widetilde{C} _t:=\widetilde{V} _t - \widetilde{G} _t.
\end{equation*}
Assume now that both $H$  and the $\{S _n\}_{n \in\{0, \ldots, T\}}$ are in $L ^2(\Omega, P)$. A generalized trading strategy is called {\bfi  admissible} for $H$ whenever it is in $L ^2(\Omega,P) $ and its associated value process is such that
\begin{equation*}
V _T=H, \quad {\rm P}\ {\it a.s.},\quad  \quad V _t, G _t  \in L ^2(\Omega,P), \quad \mbox{for each} \ t.
\end{equation*}
The hedging technique via local risk minimization consists of finding the strategies $(\widehat{\xi}^0, \widehat{\xi})$  that minimize the {\bfi  local risk process} 
\begin{equation}
\label{local risk process}
R _t(\xi^0, \xi):=E^P \left[(\widetilde{C} _{t+1}-\widetilde{C} _t)^2\mid \mathcal{F}_t\right], \quad t=0, \ldots, T-1,
\end{equation}
within the set of admissible strategies $(\xi^0, \xi)$. More specifically, 
the admissible strategy  $(\widehat{\xi}^0, \widehat{\xi})$ is called {\bfi  local risk-minimizing}  if
\begin{equation*}
R _t(\widehat{\xi}^0, \widehat{\xi})\leq R _t(\xi^0, \xi), \quad {\rm P}\ {\it a.s.}
\end{equation*}
for all $t$ and each admissible strategy $(\xi^0, \xi)$. It can be shown that~\cite[Theorem 10.9]{foellmer:schied:book} an admissible strategy is local risk-minimizing if and only if the discounted cost process is a $P$-martingale and it is strongly orthogonal to $\widetilde{S} $, in the sense that ${\rm cov} _t(\widetilde{S}_{t+1}-\widetilde{S }_t, \widetilde{C}_{t+1}-\widetilde{C} _t)=0 $, $P$-a.s., for any $t=0, \ldots, T-1 $. More explicitly, whenever a local risk-minimizing technique exists, it is fully determined by the backwards recursions:
\begin{eqnarray}
V _T &= & H,\label{hedge general 1}\\
\xi_{t+1} &= & \frac{{\rm cov}^P(\widetilde{V}_{t+1},\widetilde{S}_{t+1}-\widetilde{S }_t\mid \mathcal{F} _t)}{{\rm var}^P(\widetilde{S}_{t+1}-\widetilde{S }_t\mid \mathcal{F} _t)},\label{hedge general 2}\\
\widetilde{V} _t &= &E^P\left[\widetilde{V}_{t+1}\mid \mathcal{F} _t\right]- \xi_{t+1}E^P\left[\left(\widetilde{S}_{t+1}-\widetilde{S }_t\right)\mid \mathcal{F} _t\right].\label{hedge general 3}
\end{eqnarray}
The initial investment $V _0 $ determined by these relations will be referred to as the {\bf local risk minimization option price} associated to the measure $P$.

\subsection{Local risk minimization with respect to a martingale measure}
\label{Local risk minimization with respect to a martingale measure}

As we have explicitly indicated in expressions~(\ref{hedge general 1})--(\ref{hedge general 3}), the option values and the hedging ratios that they provide depend on a previously chosen probability measure and a filtration. In the absence of external sources of information, the natural filtration  $\{\mathcal{F}_t\}$ to be used is the one generated by the price process. Regarding the choice of the probability measure, the first candidate that should be considered is the physical or historical probability associated to the price process due to the fact that from a risk management perspective this is the natural measure that should be used in order to construct the local risk~(\ref{local risk process}). 

In practice, the use of the physical probability encounters two major difficulties: on one hand, when~(\ref{hedge general 1})--(\ref{hedge general 3}) are written down with respect to this measure, the resulting expressions are convoluted and numerically difficult to estimate due to the high variance of the associated Monte Carlo estimators; a hint of this difficulty in the GARCH situation can be seen in Proposition 2.5 of~\cite{ortega:garch:pricing}. Secondly, unless the discounted prices are martingales with respect to the physical probability measure or there is a minimal martingale measure available, the option prices that result from this technique cannot be interpreted as arbitrage free prices. 

These reasons lead us to explore the local risk minimization strategy for martingale measures and, more specifically for the martingale measures introduced in Section~\ref{Equivalent martingale and quasi-martingale measures for stochastic volatility models}. Given that the trend terms that separate the physical measure from being a martingale measure are usually very small when dealing with daily or weekly financial returns, it is expected that the inaccuracy committed by using  a conveniently chosen martingale measure will be smaller than the numerical error that we would face in the Monte Carlo evaluation of the expressions corresponding to the physical measure; see~\cite[Proposition 3.2]{ortega:garch:pricing} for an argument in this direction in the GARCH context. In any case, the next section will be dedicated to carry out an empirical comparison between the hedging performances obtained using the different measures in Section~\ref{Equivalent martingale and quasi-martingale measures for stochastic volatility models}, as well as the various volatility estimation techniques that we described and that are necessary to use them.

The following proposition is proved using a straightforward recursive argument in expressions~(\ref{hedge general 1})--(\ref{hedge general 3}) combined with the martingale hypothesis in its statement. 

\begin{proposition}
Let  $Q$ be an  equivalent martingale measure for the price process $\{ S _t\} $ and $H(S _T)$ be a European contingent claim that depends on the terminal value of the risky asset  $S _t $. The local risk minimizing strategy with respect to the measure $Q$ is determined by the recursions:
\begin{eqnarray}
V _T &= & H,\label{hedge general martingale 1}\\
\xi_{t+1} &= & \frac{1}{\Sigma_{t+1}^2}E^Q_{t}\left[e^{-(R _T+ R _t)}H(S _T)\left(S_{t+1}e^{-r_{t+1}}-S _t\right)\right],\label{hedge general martingale 2}\\
V _t &= &E^Q_{t}\left[e^{-(R _T-R _t)}H(S _T)\right],\label{hedge general martingale 3}
\end{eqnarray} 
where 
\begin{equation}
\label{sigma as variance}
\Sigma_{t+1}^2:= {\rm var} ^Q(\widetilde{S} _{t+1}-\widetilde{S} _t\mid \mathcal{F} _t) = e^{-2R _t}E^Q_{t}\left[S _{t+1}^2e^{-2 r_{t+1}}-S _t^2\right]=e^{-2R _t}{\rm var} ^Q(S _{t+1} {\rm e}^{-r_{t+1}}\mid \mathcal{F} _t).
\end{equation}
\end{proposition}

\begin{remark}
\normalfont
When expression~(\ref{hedge general martingale 3}) is evaluated at $t=0 $ it yields the initial investment $V _0 $ necessary to setup the generalized local risk minimizing trading strategy that fully replicates the derivative $H$ and coincides with the arbitrage free price for $H$ that results from using $Q$ as a pricing measure. Obviously, this connection only holds when local risk minimization is carried out with respect to a martingale measure.
\end{remark}

\begin{remark}
\normalfont
Local risk-minimizing trading strategies computed with respect to a martingale measure $Q$ also minimize~\cite[Proposition 10.34]{foellmer:schied:book} the so called {\bf  remaining conditional risk}, defined as the process $R ^Q _t (\xi^0, \xi) :=E_t[(\widetilde{C} _T-\widetilde{C} _t)^2]$, $t=0, \ldots, T $; this is in general not true outside the martingale framework (see~\cite[Proposition 3.1]{schweizer:2001} for a counterexample). Analogously,  local risk minimizing strategies are also {\bf variance-optimal}, that is, they minimize $E^Q\left[ \left(\widetilde{H}-V _0-\widetilde{G} _T\right)^2\right]$ (see~\cite[Proposition 10.37]{foellmer:schied:book}). This is particularly relevant in the ARSV context in which a standard sufficient condition (see~\cite[Theorem 10.40]{foellmer:schied:book}) that guarantees that local risk minimization with respect to the physical measure implies variance optimality does not hold; we recall that  this condition demands for the deterministic character of the quotient 
\begin{equation*}
\beta _t:=\frac{\left(E^P_{t-1}\left[S _t-S_{t-1}\right]\right)^2}{{\rm var}_{t-1}^P\left[ S _t-S_{t-1}\right]}.
\end{equation*}
Indeed, expressions~(\ref{expected increase 1}) and~(\ref{expected increase 2}), together with Lemma~\ref{equality with Ks} imply that in our situation
\begin{equation*}
\beta _t:=\frac{e^{2K_{\epsilon _t}^P(\sigma _t)}}{\left[e^{K_{\epsilon _t}^P(2\sigma _t)}-e^{2K_{\epsilon _t}^P(\sigma _t)}\right]},
\end{equation*}
which is obviously not deterministic.
\end{remark}

\begin{remark}
\label{remark about denominator}
\normalfont
The last equality in~(\ref{sigma as variance}) is of much importance when using Monte Carlo simulations to evaluate~(\ref{hedge general martingale 1})--(\ref{hedge general martingale 3}) and suggests the correct estimator that must be used in practice: if we generate under the measure $Q$ a set of $N$ price paths all of which start at $S _t $ at time $t$  and take values $S_{t+1}^1, \ldots, S_{t+1}^N $ at time $t+1 $, then in view of the last equality in~(\ref{sigma as variance}) we write
\begin{equation*}
\Sigma_{t+1}^2=e^{-2R _t}{\rm var} ^Q(S _{t+1} {\rm e}^{-r_{t+1}}\mid \mathcal{F} _t)\simeq \frac{e^{-2R _t}}{N}\sum_{i=1}^N \left(S _{t+1}^i {\rm e}^{-r_{t+1}}-S _t \right)^2.
\end{equation*}
We warn the reader that an estimator for $\Sigma_{t+1}^2 $ based on the Monte Carlo evaluation of 
\[e^{-2R _t}E^Q_{t}\left[S _{t+1}^2e^{-2 r_{t+1}}-S _t^2\right] \] would have much more variance and once inserted in the denominator of~(\ref{hedge general martingale 2}) would produce unacceptable results.
\end{remark}

\begin{remark}
\normalfont
The Monte Carlo estimation of the expressions~(\ref{hedge general martingale 1})--(\ref{hedge general martingale 3}) requires the generation of price paths with respect to the martingale measure $Q$. In the GARCH context this can be easily carried out for a variety of pricing kernels by rewriting the process in connection with the new equivalent measure in terms of new innovations (see for example~\cite{duan:GARCH:pricing, ortega:garch:pricing, chorro:guegan:ielpo, Badescu:option:pricing}). This method can be combined with modified Monte Carlo estimators that enforce the martingale condition and that have a beneficial variance reduction effect like, for example, the Empirical Martingale Simulation technique introduced in~\cite{duan:ems1, duan:ems2}. Unfortunately, this is difficult to carry out in our context and, more importantly, it does not produce good numerical results. Given that all the pricing measures introduced in the previous section are constructed out of the physical one, the best option in our setup consists of carrying out the path simulation with respect to the physical measure and  computing the expectations in~(\ref{hedge general martingale 1})--(\ref{hedge general martingale 3}) using the corresponding Radon-Nikodym derivative. More specifically,  let $Q$ be an equivalent martingale measure that is obtained out of the physical measure $P$ by constructing a Radon-Nikodym derivative of the form:
\begin{equation*}
\frac{dQ}{dP}=\prod_{k=1}^T N _k,
\end{equation*}
such that the process $\{Z _t\} $ defined by $Z _t:=\prod_{k=1}^t N _k $ is a $P$-martingale that satisfies $ E^P\left[Z _t\right]=1 $; notice that all the measures introduced in the previous section are of that form. In that setup we define the process $\{F _t\} $ by $F _t:=\prod_{i=t}^T N _i$ that allows us to rewrite $Q$-expectations in terms of $P$-expectations, that is, for any $\mathcal{F}_T$ measurable function $f$ and any $ t \in \left\{ 0, \ldots, T\right\}$,
\begin{equation*}
E^Q\left[f\mid \mathcal{F}_t\right]=E^P\left[F_{t+1}f\mid \mathcal{F}_t\right].
\end{equation*}
The proof of this equality is a straightforward consequence of~\cite[Proposition A.11]{foellmer:schied:book} and of the specific way in which the measure $Q$ is constructed.
\end{remark}

\subsection{Local risk minimization and changes in the hedging frequency}
\label{Local risk minimization and changes in the hedging frequency}

As we already pointed out one of the major advantages of the local risk minimization hedging scheme when compared to other sensitivity based methods, is its adaptability to prescribed changes in the hedging frequency. Indeed, suppose that the life of the option $H$ with maturity in $T$ time steps is partitioned into identical time intervals of duration $j$; this assumption implies the existence of an integer $k$ such that  $kj=T $.

We now want to set up a  local risk minimizing replication strategy for $H$ in which hedging is carried out once every $j$ time steps. We will denote by $\xi_{t+j} $ the hedging ratio at time $t$ that presupposes that the next hedging will take place at time $t+j $. The value of such ratios will be obtained by minimizing the $j$-spaced local risk process:
\begin{equation*}
R _t^j(\xi^0, \xi):=E^P \left[(\widetilde{C} _{t+j}^j-\widetilde{C} _t^j)^2\mid \mathcal{F}_t\right], \quad t=0, j, 2j,\ldots, (k-1)j=T-j,
\end{equation*}
where $\{\widetilde{C} _t^j \}$ is a cost process constructed out of value and gains processes, $\{V _t ^j\} $ and $\left\{ G _t ^j\right\}$ that only take into account the prices of the underlying assets at time steps $t=0, j, 2j,\ldots, kj=T $, in particular, given an integer $l$ such that $t=lj $ 
\begin{equation*}
\widetilde{G} _t^j:=\sum_{r=1}^l \xi_{rj}\cdot (\widetilde{S} _{rj}-\widetilde{S}_{(r-1)j}).
\end{equation*}
A straightforward modification of the argument in~\cite[Theorem 10.9]{foellmer:schied:book} proves that the solution of this local risk minimization problem with modified hedging frequency with respect to a martingale measure is given by the expressions: 
\begin{eqnarray}
V _T^j &= & H,\label{hedge general martingale frequency 1}\\
\xi_{t+j} &= & {\rm e}^{-(R _T-R _t)}\frac{E^Q_{t}\left[H(S _T)\left(S_{t+j}e^{-(R_{t+j}-R _t)}-S _t\right)\right]}{E^Q_{t}\left[S_{t+j}^2{\rm e}^{-2(R_{t+j}-R _t)}-S _t ^2\right]},\label{hedge general martingale frequency 2}\\
V _t ^j&= &E^Q_{t}\left[e^{-(R _T- R _t)}H(S _T)\right],\label{hedge general martingale frequency 3}
\end{eqnarray} 
for any $t=0, j, 2j,\ldots, (k-1)j=T-j $. As a follow up to what we pointed out in the remark~\ref{remark about denominator}, the denominator in~(\ref{hedge general martingale frequency 2}) should be computed by first noticing that
\begin{equation*}
E^Q_{t}\left[S_{t+j}^2{\rm e}^{-2(R_{t+j}-R _t)}-S _t ^2\right]= {\rm var} _t^Q \left[
S_{t+j}{\rm e}^{-(R_{t+j}-R _t)}\right],
\end{equation*}
and then using the appropriate Monte Carlo estimator for the variance, that is,
\begin{equation}
\label{variance for denominator general}
{\rm var} ^Q(S _{t+1} {\rm e}^{-r_{t+1}}\mid \mathcal{F} _t)\simeq \frac{1}{N}\sum_{i=1}^N \left(S _{t+j}^i {\rm e}^{-(R_{t+j}^i-R _t)}-S _t \right)^2,
\end{equation}
where $S_{t+j}^1, \ldots, S_{t+j}^N $ are $N$ realizations of the price process under $Q$ at time $t+j $, using paths that have all the same origin $S _t$ at time $t$. If the interest rate process $\left\{ r _t\right\}$ is constant and equal to $r$, then~(\ref{variance for denominator general}) obviously reduces to
\begin{equation*}
{\rm var} ^Q(S _{t+1} {\rm e}^{-r_{t+1}}\mid \mathcal{F} _t)\simeq \frac{1}{N}\sum_{i=1}^N \left(S _{t+j}^i {\rm e}^{-jr}-S _t \right)^2.
\end{equation*}

\subsection{Local risk minimization and the minimal martingale measure}

Local risk-minimization  requires picking a particular probability measure in the problem. We also saw that it is only when discounted prices are martingales with respect to that measure that the obtained value process for the option in question coincides with the arbitrage free price that one would obtain by using that martingale measure as a pricing kernel. This observation particularly concerns the physical probability which one would take as the first candidate to use this technique since it is the natural measure to be used when quantifying the local risk.

The solution of this problem is one of the motivations for the introduction of the minimal martingale measure $Q_{{\rm min}}$ that we defined in Section~\ref{Equivalent martingale and quasi-martingale measures for stochastic volatility models}. Indeed, it can be proved that (see Theorem 10.22 in~\cite{foellmer:schied:book}) the value process of the local risk-minimizing strategy {\it with respect to the physical measure} coincides with arbitrage free price for $H$ obtained by using $Q_{{\rm min}}$ as a pricing kernel. More specifically, if $V _t ^P $ is the value process for $H$ obtained out of formulas~(\ref{hedge general 1})--(\ref{hedge general 3}) using the physical measure then 
\begin{equation}
\label{values with minimal martingale}
V _t ^P=E^{Q_{{\rm min}}}_{t}\left[e^{-(R _T- R _t)}H(S _T)\right].
\end{equation}

We emphasize that, as we pointed out in Remark~\ref{minimal is signed}, the minimal martingale measure in the ARSV setup is in general signed. Nevertheless, the occurrences of negative Radon-Nikodym derivatives are extremely unlikely, at least when dealing with Gaussian innovations, which justifies its use for local risk minimization.

We also underline that even though the value processes obtained when carrying out local risk minimization with respect to the physical and the minimal martingale measures are identical, the hedges are in general {\it not} the same and consequently so are the hedging errors. In the next proposition we provide a relation that links both hedges and identify situations in which they coincide. Before we proceed we need to introduce the notion of {\bf  global (hedging) risk process}: it can be proved that once a probability measure $P$ has been fixed, if there exists a local risk-minimizing strategy  $(\xi^0, \xi)$ with respect to it, then it is unique (see~\cite[Proposition 10.9]{foellmer:schied:book}) and the discounted payoff $\widetilde{H}$ can be decomposed as (see~\cite[Corollary 10.14]{foellmer:schied:book})
\begin{equation}
\label{decomposition payoff risk minimizing}
\widetilde{H}=V _0+\widetilde{G} _T+\widetilde{L} _T,
\end{equation}
with $\widetilde{G} _t $ the discounted gains process associated to $(\xi^0, \xi)$ and $\widetilde{L} _t:=\widetilde{C} _t-C _0 $, $t=0, \ldots,T $ a sequence $\{ \widetilde{L} _t\}_{t \in \{0, \ldots, T\}}$ that we will call the (discounted) global (hedging) risk process. $\{\widetilde{L} _t\}_{t \in \{0, \ldots, T\}}$ is a square integrable $P$-martingale that satisfies $L _0 =0 $ and that is {\bf  strongly orthogonal} to $\widetilde{S}$ in the sense that  
\begin{equation*}
{\rm cov} ^P((\widetilde{L}_{t+1}-\widetilde{L} _t)(\widetilde{S}_{t+1}-\widetilde{S} _t)\mid \mathcal{F} _t)=0 \quad \mbox{for any} \quad t=0, \ldots, T-1.
\end{equation*}
The decomposition~(\ref{decomposition payoff risk minimizing}) shows that $\widetilde{L} _T $ measures how far $\widetilde{H} $ is from the terminal value of the self-financing portfolio uniquely determined by the initial investment $V _0 $ and the trading strategy given by $\{\xi _t\} $ (see~\cite[Proposition 1.1.3]{lamberton:lapeyre}).

\begin{proposition}
\label{hedges under change of measure}
Let $H(S _T)$ be a European contingent claim that depends on the terminal value of the risky asset  $S _t $ and let $\{ \xi _t^{Q_{{\rm min}}}\}_{t=1}^T $ and $\{ \xi _t^{P}\}_{t=1}^T $ be the local risk minimizing hedges associated to the minimal martingale measure $Q_{{\rm min}} $ and the physical measure $P$, respectively. Let $\left\{ \widetilde{L }_t ^P\right\}_{t=0}^T $ be the associated $P$-global risk process. Then, for any $t \in \left\{ 1, \ldots, T\right\}$:
\begin{equation}
\label{relation hedges}
\xi_t^{Q_{{\rm min}}}= \xi _t ^P+ \frac{E^{Q_{{\rm min}}}_{t-1}\left[ \widetilde{L }_t ^P (\widetilde{S} _t- \widetilde{S}_{t-1})\right]}{{\rm var}_{t-1}^{Q_{{\rm min}}} \left[\widetilde{S} _t- \widetilde{S}_{t-1}\right]}.
\end{equation} 
If the processes $\left\{ \widetilde{L }_t ^P\right\}_{t=0}^T $ and $\left\{ \widetilde{S}_t\right\}_{t=0}^T $ are either $ Q_{{\rm min}} $--strongly orthogonal or $Q_{{\rm min}} $--independent, then $\{ \xi _t^{Q_{{\rm min}}}\}_{t=1}^T =\{ \xi _t^{P}\}_{t=1}^T $.
\end{proposition}

\section{Empirical study}
\label{Empirical study}

The goal of this section is comparing the hedging performances obtained by implementing the local risk minimization scheme using the different volatility estimation techniques and martingale measures introduced in Section~\ref{Auto-regressive stochastic volatility (ARSV) models}. In order to enrich the discussion, we will add to the list two standard sensitivity based hedging methods that provide good results in other contexts, namely standard Black-Scholes~\cite{black:scholes:paper} and an adaptation of Duan's static hedge~\cite{duan:GARCH:pricing} to the ARSV context.

The way in which we proceed with the comparison consists of  taking a particular ARSV model as price paths generating process and then carrying out the replication of a European call option according to the prescribed list of hedging methods for various moneyness, strikes and hedging frequencies. At the end of the life of each of this option, the square hedging error is recorded. This task is carried out for a number of independently generated price paths which allows us to estimate a mean square hedging error that we will use to compare the different hedging techniques.

\medskip

\noindent {\bf The price generating ARSV process.} The price process chosen for our simulations is obtained by taking the model that has as associated log-returns one of the ARSV prescriptions studied in~\cite{delcastillo:lee:2}. It consists of taking the following parameters in~(\ref{arsv model}):
\begin{equation*}
r=0.1/252, \qquad \gamma=-0.821, \qquad \phi=0.9, \qquad \sigma _w=0.675.
\end{equation*}
The use of formulas~(\ref{variance and kurtosis}) shows that the resulting log-returns are very leptokurtic (${\rm kurtosis}=33.00 $) and volatile (the stationary variance is $9.01 \cdot 10^{-4} $ which amounts to an annualized volatility of 47\%). The intentionality behind this choice is exacerbating the defects of the different hedging methods in order to better compare them.

\medskip

\noindent {\bf The hedging methods.} Given the price paths generated by the ARSV process detailed in the previous paragraph, we will carry out the replication of European call options that have them as underlying asset using local risk minimization with respect to the two martingale measures introduced in Section~\ref{Equivalent martingale and quasi-martingale measures for stochastic volatility models}, namely:
\begin{itemize}
\item {\bf ARSV mean correcting martingale measure introduced in Theorem~\ref{quasi mean correcting martingale measure}}. The hedging method is abbreviated as LRM-MCMM in the figures. 
\item {\bf Minimal martingale measure}. Denoted by $Q_{{\rm min}}$ and introduced in Theorem~\ref{introduction minimal martingale measure}. The hedging method is abbreviated as LRM-MMM in the figures. This measure may be signed in the ARSV context but this happens with extremely low probability (see Remark~\ref{minimal is signed}); in our simulations we did not encounter this situation even a single time. Local risk minimization with respect to this measure yields the same value process as the physical probability but not the same hedging ratios; Proposition~\ref{hedges under change of measure} spells out the difference and provides sufficient conditions for them to coincide.
\end{itemize}
The construction of all these measures requires conditional expectations of functions of the conditional variance  of the log-returns with respect to the filtration generated by the price process. As this filtration does not determine the conditional volatilities, those expectations cannot be easily computed; we hence proceed by estimating the conditional volatilities via 
Kalman filtering and a prediction-update routine based on the $h$-likelihood approach, both briefly described in Section~\ref{Volatility and model estimation}. Different volatility estimation techniques lead to different pricing/hedging kernels and hence to different hedging performances that we compare in our simulations.

We will complete the comparison exercise with two more methods:
\begin{itemize}
\item {\bf Black-Scholes}: we forget that the price process is generated via an ARSV model and we handle the hedging using the standard Black-Scholes delta~\cite{black:scholes:paper}  as if the underlying was a realization of a log-normal process with constant drift and volatility given by the drift and marginal volatility (given by the expression~(\ref{variance and kurtosis})) of the ARSV model specified above.
\item {\bf Duan's static delta hedge}: this is a generalization of the Black-Scholes delta hedge that has been introduced in~\cite[Corollary 2.4]{duan:GARCH:pricing} in the GARCH context. In that result the author computes the derivative of a European call price (the delta put can be readily obtained out of the call-put parity) with respect to the price of the underlying in an arbitrary incomplete situation in which that option price is simply obtained as the expectation of the discounted payoff with respect to a given pricing kernel $Q $, that is:
\begin{equation*}
V _t = E^Q\left[e^{-(R _T-R _t)}H(S _T)\mid \mathcal{F}_t\right].
\end{equation*}
In those circumstances:
\begin{equation*}
\xi^{{\rm Duan}}_{t+1}:=\frac{\partial V _t}{\partial S _t}=e^{-(R _T-R _t)} E^Q\left[\frac{S _T}{S _t}\boldsymbol{1}_{S _T\geq K}\mid \mathcal{F} _t\right].
\end{equation*}
In our empirical study we will use this hedging technique using the various pricing kernels previously introduced. A {\it dynamical} refinement of this hedge can be obtained by considering additional dependences of the value process on the underlying price that may occur (like in our case) through the volatility process~\cite{garcia:renault, Badescu:Ortega}. This will not be considered in our study.
\end{itemize}

\medskip

\noindent {\bf Numerical specifications.} The comparisons in performance will be carried out by hedging batches of European call options with different maturities and three different moneyness $S _0/K $ of 1.11, 1, and 0.90, and for a number of independent price paths. The initial price $S _0$ of the underlying is always equal to $100 $. We will assign to each hedging experiment a prescribed hedging frequency. The mean square hedging error associated to each individual hedging method will be computed by taking the mean of the terminal square hedging errors for each of the price paths. In the particular case of the local risk minimization method, where the hedges are obtained by computing the expressions~(\ref{hedge general martingale frequency 1})--(\ref{hedge general martingale frequency 3}), the conditional expectations that they contain will be estimated via Monte Carlo evaluations based on the use of $2500 $ price paths.

\medskip

\noindent {\bf The hedging exercises.} We now explain in detail the different hedging experiments that we have carried out.
\begin{itemize}
\item {\bf Exercise 1}: all the hedging techniques explained in the paper are used. The four maturities considered are $6$, $8$, $10 $, and $12 $ time steps. Hedging is carried out daily. The computation of the mean square hedging error is based on $1000 $ price paths. The first thing that we notice is that, even for these short maturities, the performance of Duan's static hedge is very poor. This is specific to the ARSV situation for in the GARCH case this technique provides very competitive results~\cite{ortega:garch:pricing} even in the presence of processes excited by leptokurtic innovations~\cite{Badescu:Ortega}. Additionally, the Black-Scholes hedging scheme provides remarkably good results, specially  for in and at the money options. Regarding  the local risk minimization strategy, the best results are obtained when using a combination of a Kalman based estimation of the conditional volatility and the minimal martingale measure $Q_{{\rm min}}$. 

\begin{figure}[!htp]
\includegraphics[scale=.25,angle=-90]{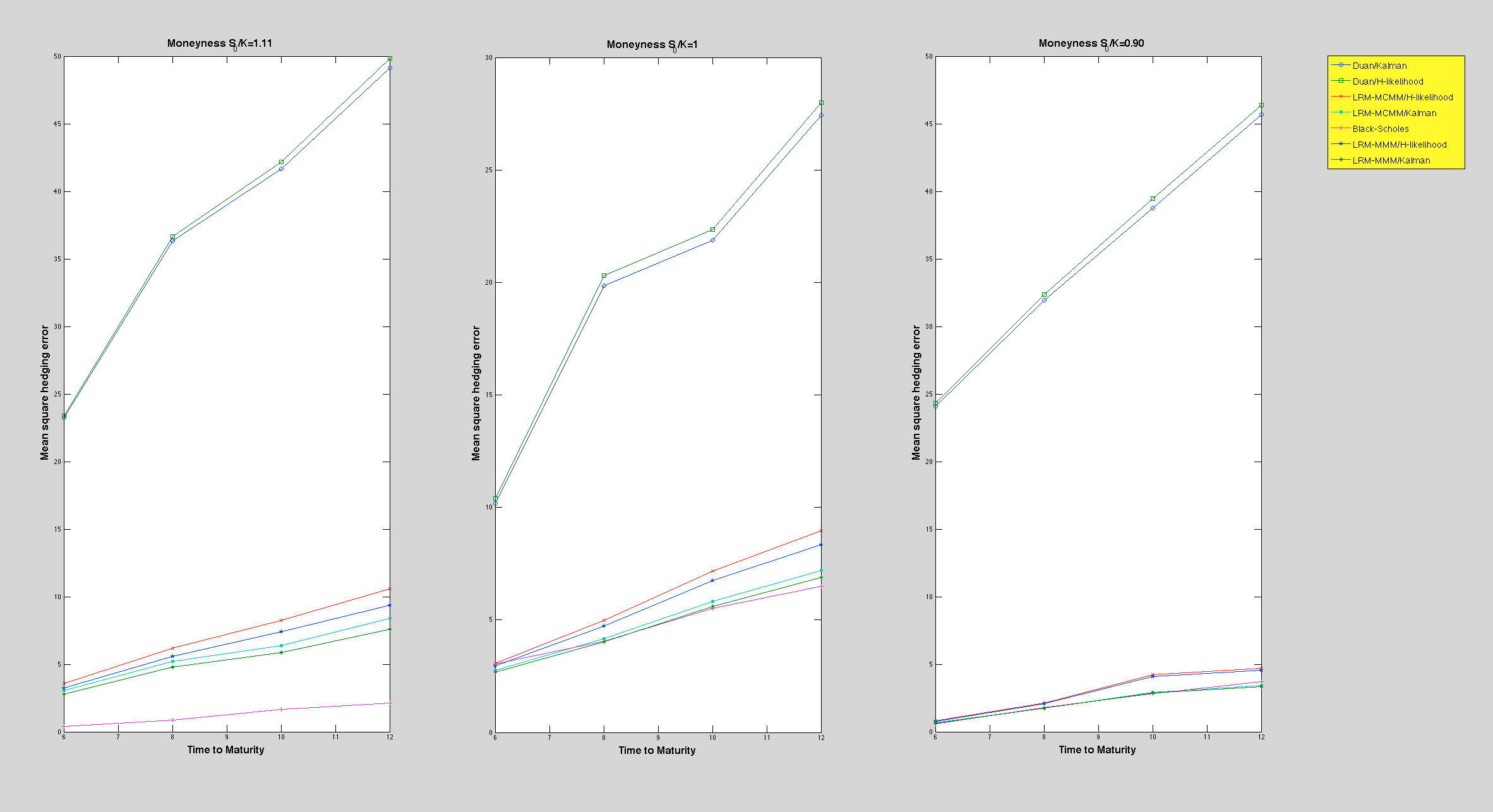}
\caption{Experiment 1. All the hedging techniques explained in the paper are used. The four maturities considered are $6$, $8$, $10 $, and $12 $ time steps. Hedging is carried out daily. The computation of the mean square hedging error is based on $1000 $ price paths.}
\label{fig:prices}
\end{figure}

\item {\bf Exercise 2}: we have retaken the specifications of Exercise $1 $, but this time around we have eliminated from the list of techniques under examination Duan's static hedge due to its poor performance. The four maturities considered are $10$, $20$, $30 $, and $40 $ time steps. Hedging is carried out every $10 $ days. The main difference with the previous exercise is the degradation of the performance of the Black-Scholes scheme in comparison with local risk minimization. This is due to the fact that, unlike Black-Scholes, local risk minimization admits adjustment to the change of hedging frequency by using the formulas~(\ref{hedge general martingale frequency 1})--(\ref{hedge general martingale frequency 3}). Regarding local risk minimization it is still the combination of the minimal martingale measure with Kalman based estimation for the volatilities that performs the best.

\begin{figure}[!htp]
\includegraphics[scale=.25,angle=-90]{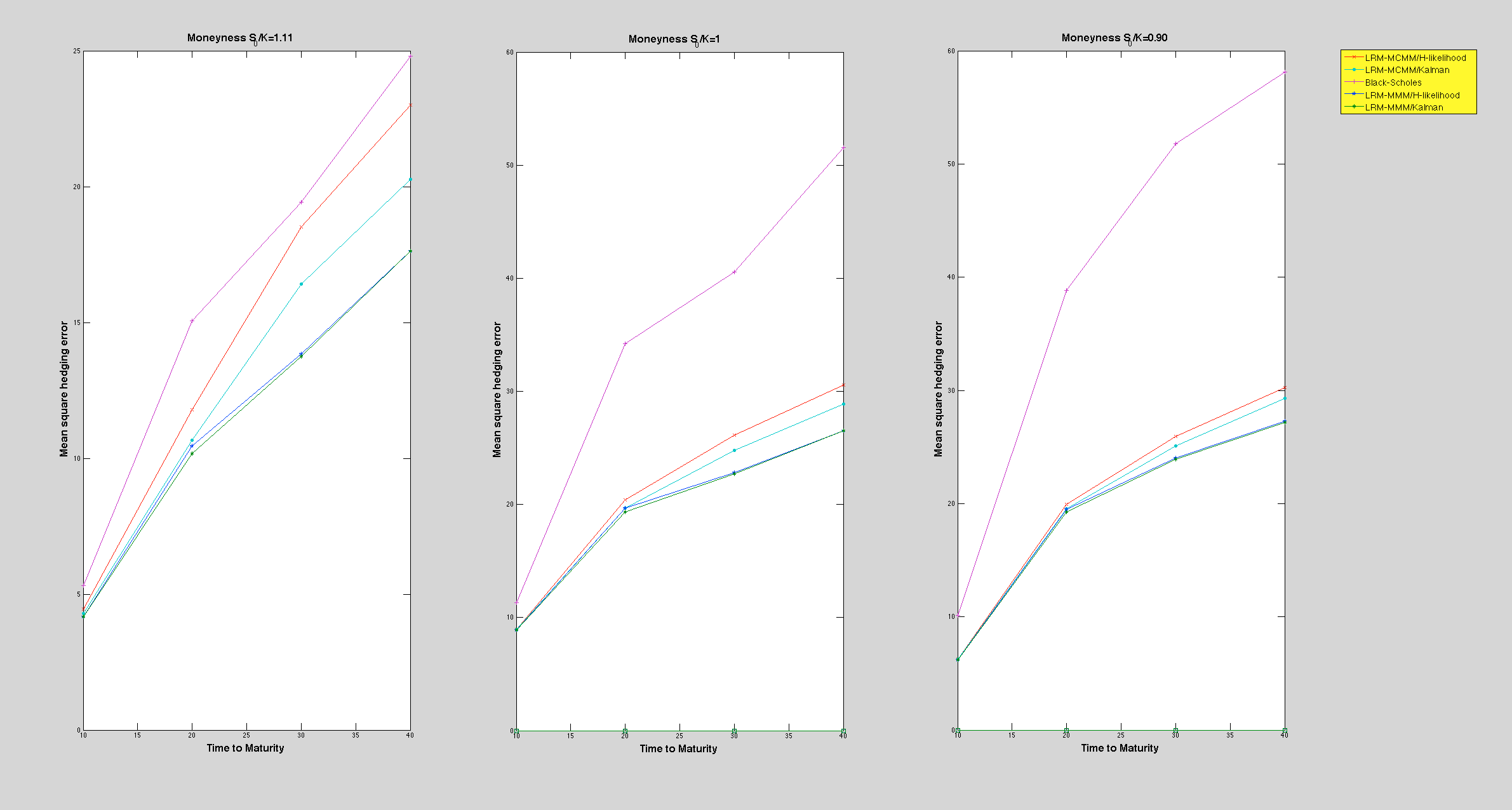}
\caption{Experiment 2. The four maturities considered are $10$, $20$, $30 $, and $40 $ time steps. Hedging is carried out every $10 $ days. The computation of the mean square hedging error is based on $1000 $ price paths.}
\label{fig:prices}
\end{figure}

\item {\bf Exercise 3}: we experiment with longer maturities and smaller hedging frequencies. We consider maturities of $20 $, $40 $, $60 $, $80 $, $100 $,  and $120 $ time steps with hedging carried out every $20 $ days. The computation of the mean square hedging error is based on $600 $ price paths. The novelty in these experiments with respect to the preceding ones is that in some instances the volatility estimation via $h$--likelihood seems more appropriate than Kalman's method in this setup.

\begin{figure}[!htp]
\includegraphics[scale=.25,angle=-90]{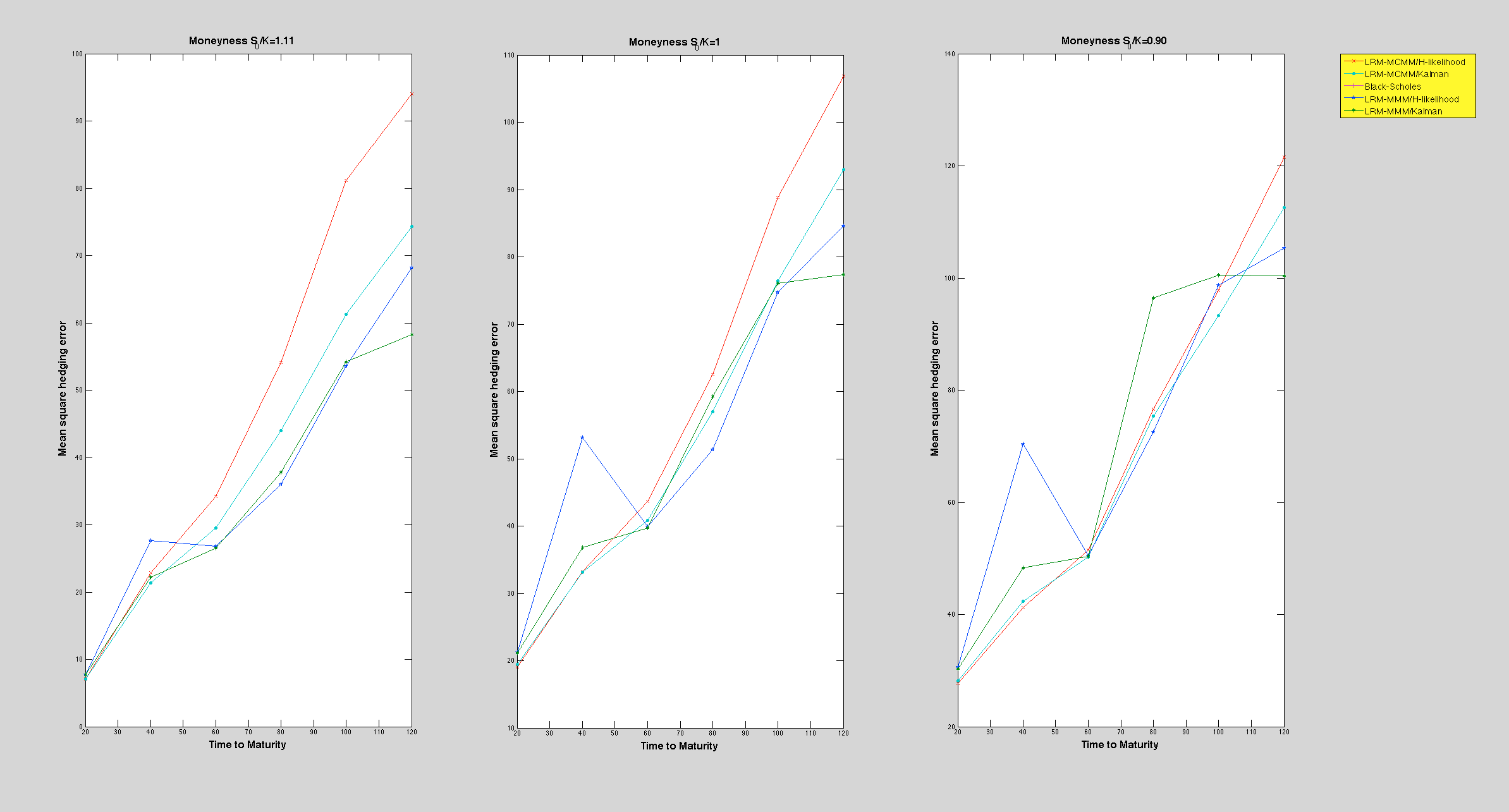}
\caption{Experiment 6.  The hedging techniques used are spelled out in the legend. Six maturities are considered: $40$, $60$, $80 $, and $100$ time steps. Hedging is carried out every $20 $ days. The computation of the mean square hedging error is based on $600 $ price paths.}
\label{fig:prices}
\end{figure}

\end{itemize}

\section{Conclusions}

In this work we have reported on the applicability of the local risk minimization pricing/hedging technique in the handling of European options that have an auto-regressive stochastic volatility (ARSV) model as their underlying asset. The method has been implemented using the following martingale measures: 
\begin{itemize}
\item The minimal martingale measure: even though it is in general signed, the probability of occurrence of negative Radon-Nikodym derivatives is extremely low and hence can be used to serve our purposes.
\item Mean correcting martingale measure: it is a measure that shares the same expression and good analytical and numerical properties as the measure obtained in the GARCH context out of the Extended Girsanov Principle. The reason to use this proxy instead of the genuine Extended Girsanov Principle is that in the ARSV case, this procedure produces a measure that is numerically expensive to evaluate. 
\end{itemize}
An added difficulty  when putting this scheme into practice and that is specific to ARSV models, has to do with the need of estimating the volatility that, even though in this situation is not determined by the observable price process, it is necessary in order to construct the pricing/hedging kernels. Two techniques are used to achieve that: Kalman filtering and hierarchical likelihood ($h$-likelihood). Both methods are adequate  volatility estimation techniques, however the $h$-likelihood technique has a much wider range of applicability for it is not subjected to the rigidity of the state space representation necessary for Kalman and hence can be used for  stochastic volatility models with  complex link functions or when innovations are non-Gaussian. 

We have carried out a numerical study to compare the hedging performance of the methods based on different measures and volatility estimation techniques. Overall, the most appropriate technique seems to be the choice that combines the minimal martingale measure with a Kalman filtering based estimation of the volatility. $h$-likelihood based volatility estimation gains pertinence as maturities become longer and the hedging frequency diminishes.

\section{Appendix}
\label{Appendix}

\subsection{Proof of Lemma~\ref{relation cumulant functions}}

Let  $Z$ be an arbitrary $\mathcal{F}_{t-1} $ measurable random variable. As $ \epsilon _t $ is independent of $\mathcal{F}_{t-1} $ and by hypothesis it is also independent of $u$, the joint law $P_{\epsilon _t, u,Z} $ of the random variable $( \epsilon _t, u,Z):(\Omega,P)\rightarrow \mathbb{R} ^3 $ can be written as the product $P_{\epsilon _t, u,Z} (x,y,z)=P_{\epsilon_t} (x)P _{u,Z} (y,z)$. Hence, using Fubini's Theorem we have
\begin{eqnarray}
E^P\left[e^{u \epsilon _t}Z\right]&=&\int\int e^{yx}z dP_{\epsilon_t} (x)dP _{u,Z} (y,z)
=\int\left[\int e^{yx} dP_{\epsilon_t} (x)\right]z dP _{u,Z} (y,z)\notag\\
	&= &\int e^{L_{\epsilon _t}(y)}z dP _{u,Z} (y,z)= E^P\left[e^{L_{\epsilon _t}(u)}Z\right].\label{intermediate conditional}
\end{eqnarray}
At the same time, by the definition of $K_{\epsilon _t}^P$, for any $\mathcal{F}_{t-1} $ measurable random variable $Z$, the equality~(\ref{intermediate conditional}) implies that:
\begin{equation*}
E^P\left[e^{K_{\epsilon _t}^P (u)}Z\right]=E^P\left[e^{u\epsilon _t}Z\right]=E^P\left[e^{L_{\epsilon _t}(u)}Z\right].
\end{equation*}
As $e^{K_{\epsilon _t}^P (u)} $ is $\mathcal{F}_{t-1} $ measurable and $Z$ arbitrary, the almost sure uniqueness of the conditional expectation  implies that
\begin{equation*}
e^{K_{\epsilon _t}^P (u) }=E^P\left[e^{L_{\epsilon _t}(u)}\mid \mathcal{F}_{t-1}\right],
\end{equation*}
as required. \quad $\blacksquare$

\medskip

The result that we just proved is a particular case of the following Lemma that we state for reference. The proof follows the same pattern as in the previous paragraphs.

\begin{lemma}
\label{generalized cumulant functions}
Let $\left(\Omega, \mathcal{F}, P\right) $ be a probability space and $\mathcal{B} \subset \mathcal{F} $ a sub-$\sigma$-algebra. Let $X$ and $Y$ be two random variables such that  $X$ is simultaneously independent of $Y$ and $\mathcal{B} $. For any measurable function $\Phi: \mathbb{R}^2 \rightarrow \mathbb{R} $ define the mapping $F: \mathbb{R} \rightarrow \mathbb{R} $ by $F (y)= E\left[\Phi(X,y)\right]$. Then,
\begin{equation*}
E\left[\Phi(X,Y)\mid \mathcal{B}\right]=E[F (Y)\mid \mathcal{B}].
\end{equation*}
\end{lemma}

\subsection{Proof of Theorem~\ref{mcmm arsv expression}}

The proof is a straightforward consequence of Theorem 3.1 in~\cite{elliott:madan:mcmm}. Indeed, it suffices to identify in the ARSV case the two main ingredients necessary to carry out the Extended Girsanov Principle, namely the one period excess discounted return $\mu_t $ defined by:
\begin{equation*}
e^{\mu_t}:= E^P_{t-1}\left[\frac{\widetilde{S} _t}{\widetilde{S} _{t-1}}\right]=e^{m _t-r}E^P_{t-1}\left[e^{\sigma_t \epsilon _t}\right]= \frac{e^{m _t-r}}{K},
\end{equation*}
with $K:=1/E^P_{t-1}\left[e^{\sigma_t \epsilon _t}\right] $, and the process $\{W _t\}$ uniquely determined by the relation:
\begin{equation*}
W _t:= \frac{\widetilde{S} _t}{\widetilde{S}_{t-1}}e^{-\mu _t}=Ke^{\sigma_t \epsilon _t}.
\end{equation*}
The Theorem follows from expression (3.8) in~\cite{elliott:madan:mcmm}. \quad $\blacksquare$

\subsection{Proof of Theorem~\ref{quasi mean correcting martingale measure}}

\noindent\textbf{(i)} We start by noticing that  since $\epsilon_t $ is independent of both   $ \rho_t $ and $\mathcal{F}_{t-1} $, by Lemma~\ref{generalized cumulant functions},
\begin{equation}
\label{for later 33}
E^P_{t-1}\left[N _t (\epsilon _t,  \rho _t  )\right]= E^P_{t-1}\left[F(\rho _t  )\right],
\end{equation}
where the real function $F: \mathbb{R} \rightarrow  \mathbb{R}$ is defined by
\begin{equation*}
F (y)= E^P\left[N _t (\epsilon _t,  y)\right]=\int \frac{f ^P(x+ y)}{f ^P(x)}f ^P(x) dx=\int f ^P(x+ y) dx=\int f ^P(x) dx=1,
\end{equation*}
which substituted in~(\ref{for later 33}) yields
\begin{equation}
\label{integral equal to 1}
E^P_{t-1}\left[N _t (\epsilon _t,  \rho _t  )\right]=1.
\end{equation}
This equality proves immediately the martingale property for $\{Z _t \} $. Indeed,
\begin{equation*}
E^P_{t-1}\left[Z _t \right]=E^P_{t-1}\left[N_t \right]Z  _{t-1} =Z  _{t-1}.
\end{equation*}
Finally, as $\{Z _t \} $ is a $P$-martingale
\begin{equation*}
E^P\left[Z _t  \right]= E^P\left[Z _1 \right]= E^P\left[N _1  \right]=1.
\end{equation*}

\medskip

\noindent {\bf (ii)} $Z _T  $ is by construction non-negative and $E[Z _T ]=\mathbb{P}(Z _T >0)=1 $. This guarantees  (see, for example, Remarks after Theorem 4.2.1 in~\cite{lamberton:lapeyre}) that $Q  _{{\rm mc}}$ is a probability measure equivalent to $P$.

We now start proving that $Q  _{{\rm mc}}$ is a martingale measure by showing that:
\begin{equation}
\label{simple return for martingale 1}
E^{Q^1 _{{\rm mc}}}_{t-1}\left[\frac{S _t}{S _{t-1}}\right]= E^P_{t-1}\left[e^{m _t+ \sigma _t \epsilon _t}N _t\right].
\end{equation}
Indeed,
\begin{eqnarray*}
E^{Q^1 _{{\rm mc}}}_{t-1}\left[\frac{S _t}{S _{t-1}}\right]&= & E^{Q^1 _{{\rm mc}}}_{t-1}\left[e^{m _t+ \sigma _t \epsilon _t}\right]= \frac{1}{E^P_{t-1}\left[Z _T\right]}E^P_{t-1}\left[Z _Te^{m _t+ \sigma _t \epsilon _t}\right]= \frac{1}{Z _{t-1}}E^P_{t-1}\left[Z _Te^{m _t+ \sigma _t \epsilon _t}\right]\\
	&=&
	 \frac{1}{Z _{t-1}}E^P_{t-1}\left[E^P_{T-1}\left[Z _Te^{m _t+ \sigma _t \epsilon _t}\right]\right]=  \frac{1}{Z _{t-1}}E^P_{t-1}\left[E^P_{T-1}\left[Z _T\right]e^{m _t+ \sigma _t \epsilon _t}\right]\\
	 &=&\frac{1}{Z _{t-1}}E^P_{t-1}\left[Z _{T-1}e^{m _t+ \sigma _t \epsilon _t}\right]= \cdots
	 =\frac{1}{Z _{t-1}}E^P_{t-1}\left[Z _{t}e^{m _t+ \sigma _t \epsilon _t}\right]\\
	 &= &\frac{1}{Z _{t-1}}E^P_{t-1}\left[N _tZ _{t-1}e^{m _t+ \sigma _t \epsilon _t}\right]=E^P_{t-1}\left[e^{m _t+ \sigma _t \epsilon _t}N _t\right].
\end{eqnarray*}
Now, as $\epsilon _t $ is independent of both $\sigma_t $ and $\mathcal{F} _{t-1} $, we can use Lemma~\ref{generalized cumulant functions} to prove that 
\begin{equation}
\label{second intermediate simple 1}
E^P_{t-1}\left[e^{m _t+ \sigma _t \epsilon _t}N _t\right] =  E^P_{t-1}\left[e^{m _t- \sigma _t\rho _t+L_{\epsilon _t}(\sigma _t)}\right].
\end{equation}
Indeed, we can write
\begin{eqnarray}
\label{second intermediate simple 2}
E^P_{t-1}\left[e^{m _t+ \sigma _t \epsilon _t}N _t\right] &= &e^{m _t} E^P_{t-1}\left[e^{\sigma _t \epsilon _t}N _t(\epsilon _t, \rho _t)\right]=e^{m _t} E^P_{t-1}\left[F(\sigma _t, \rho  _t)\right],
\end{eqnarray}
where the function $F: \mathbb{R} ^2\rightarrow \mathbb{R} $ is defined by
\begin{eqnarray*}
F(y,z)&=& E^P\left[e^{y \epsilon _t}N _t(\epsilon _t, z)\right]=\int e^{yx} f ^P(x) \frac{f ^P(x+z)}{f ^P(x)}d x=\int e^{y(s-z)}f ^P(s) d s\\
	&= & E^P\left[e^{y(\epsilon _t -z)}\right]= e^{-yz}e^{L_{\epsilon _t}^P (y)},
\end{eqnarray*}
which substituted in~(\ref{second intermediate simple 2}) yields~(\ref{second intermediate simple 1}). Finally, if we insert in~(\ref{second intermediate simple 1}) the explicit expression~(\ref{market prices of risk}) that defines the market price of risk, we obtain that
\begin{equation*}
E^{Q^1 _{{\rm mc}}}_{t-1}\left[\frac{S _t}{S _{t-1}}\right]=E^P_{t-1}\left[e^{m _t- \sigma _t\rho _t+L_{\epsilon _t}(\sigma _t)}\right]= e ^{r-K_{\epsilon _t}^P(\sigma _t)}E^P_{t-1}\left[ e^{L^P_{\epsilon _t}(\sigma_t)}\right]=e ^r,
\end{equation*}
as required. The last equality in the previous expression follows from Lemma~\ref{relation cumulant functions}. \quad $\blacksquare$

\subsection{Proof of Proposition~\ref{hedges under change of measure}}

Consider the unique decomposition~(\ref{decomposition payoff risk minimizing}) of the discounted payoff $\widetilde{H } $ as a sum of the discounted $P$-gains and $P$-risk processes:
\begin{equation}
\label{decomposition at T}
 \widetilde{H}=V _0+\sum_{k=1}^T \xi _k^P \cdot \left(\widetilde{S} _k- \widetilde{S}_{k-1}\right)+\widetilde{L} _T^P.
\end{equation}
We recall that the minimal martingale measure is characterized by the property that every $P$-martingale  that is strongly orthogonal to the discounted price process $\left\{ \widetilde{S}_t\right\}_{t=0}^T $, is also a $Q_{{\rm min}}$-martingale. Hence, as the $P$-martingale $\left\{ \widetilde{L }_t ^P\right\}_{t=0}^T $ is $P$-strongly orthogonal to $\left\{ \widetilde{S}_t\right\}_{t=0}^T $, it is therefore a $Q_{{\rm min}}$-martingale. Having this in mind, as well as~(\ref{values with minimal martingale}), we take conditional expectations $E^{ Q_{{\rm min}}}_{t-1}$ on both sides of the decomposition~(\ref{decomposition at T}) and we obtain:
\begin{equation*}
\widetilde{V} _t^P=V _0+\sum_{k=1}^t \xi _k^P \cdot \left(\widetilde{S} _k- \widetilde{S}_{k-1}\right)+\widetilde{L} _t^P.
\end{equation*}
As $\widetilde{V} _t^P $ is the local risk minimizing value process for both the $P$ and the $Q_{{\rm min}} $ measures, that is $\widetilde{V} _t^P= \widetilde{V} _t^{Q_{{\rm min}}}$, if we multiply both sides of by $ \left(\widetilde{S} _t- \widetilde{S}_{t-1}\right)$ and take conditional expectations $E^{ Q_{{\rm min}}}_{t-1}$ we have:
\begin{eqnarray}
E^{ Q_{{\rm min}}}_{t-1}\left[\widetilde{V} _t^{ Q_{{\rm min}}} \left(\widetilde{S} _t- \widetilde{S}_{t-1}\right)\right]&=&E^{ Q_{{\rm min}}}_{t-1}\left[V _0\left(\widetilde{S} _t- \widetilde{S}_{t-1}\right)\right]+\sum_{k=1}^t E^{ Q_{{\rm min}}}_{t-1}\left[\xi _k^P \cdot \left(\widetilde{S} _k- \widetilde{S}_{k-1}\right)\left(\widetilde{S} _t- \widetilde{S}_{t-1}\right)\right]\notag\\
	& &+E^{ Q_{{\rm min}}}_{t-1}\left[\widetilde{L} _t^P\left(\widetilde{S} _t- \widetilde{S}_{t-1}\right)\right]\notag\\
	&= &\xi _t^P E^{ Q_{{\rm min}}}_{t-1}\left[ \left(\widetilde{S} _t- \widetilde{S}_{t-1}\right)^2\right]+E^{ Q_{{\rm min}}}_{t-1}\left[\widetilde{L} _t^P \left(\widetilde{S} _t- \widetilde{S}_{t-1}\right)\right]\notag.
\end{eqnarray}
If we divide both sides of this equality by $E^{ Q_{{\rm min}}}_{t-1}\left[ \left(\widetilde{S} _t- \widetilde{S}_{t-1}\right)^2\right] $ and we use the relation~(\ref{hedge general 2}) we obtain that
\begin{equation*}
\xi_t^{Q_{{\rm min}}}= \xi _t ^P+ \frac{E^{Q_{{\rm min}}}_{t-1}\left[ \widetilde{L }_t ^P (\widetilde{S} _t- \widetilde{S}_{t-1})\right]}{{\rm var}_{t-1}^{Q_{{\rm min}}} \left[\widetilde{S} _t- \widetilde{S}_{t-1}\right]}, 
\end{equation*}
as required. Finally, we notice that
\begin{equation*}
E^{Q_{{\rm min}}}_{t-1}\left[ \left(\widetilde{L }_t ^P-\widetilde{L }_{t-1} ^P \right) (\widetilde{S} _t- \widetilde{S}_{t-1})\right]=E^{Q_{{\rm min}}}_{t-1}\left[ \widetilde{L }_t ^P (\widetilde{S} _t- \widetilde{S}_{t-1})\right]-E^{Q_{{\rm min}}}_{t-1}\left[ \widetilde{L }_{t-1} ^P (\widetilde{S} _t- \widetilde{S}_{t-1})\right]=E^{Q_{{\rm min}}}_{t-1}\left[ \widetilde{L }_t ^P (\widetilde{S} _t- \widetilde{S}_{t-1})\right],
\end{equation*}
hence, either the $Q_{{\rm min}} $-independence or the orthogonality of $\left\{ \widetilde{L }_t ^P\right\}_{t=0}^T $ and $\left\{ \widetilde{S}_t\right\}_{t=0}^T $ imply that $E^{Q_{{\rm min}}}_{t-1}\left[ \widetilde{L }_t ^P (\widetilde{S} _t- \widetilde{S}_{t-1})\right]=0 $ and hence $\{ \xi _t^{Q_{{\rm min}}}\}_{t=1}^T =\{ \xi _t^{P}\}_{t=1}^T $. \quad $\blacksquare$

\addcontentsline{toc}{section}{Bibliography}
\bibliographystyle{alpha}
\bibliography{/Users/JP17/JPO_synch/BiblioData/bibliography_econometry}
\footnotesize

\end{document}